\documentclass[aps,twocolumn,showpacs,floatfix]{revtex4}
\usepackage{amsmath,amsfonts,amssymb,graphicx,color,bbm,psfrag,dsfont}
\usepackage[english]{babel}
\usepackage[T1]{fontenc}

\newcommand{\ket} [1] {\vert #1 \rangle}
\newcommand{\bra} [1] {\langle #1 \vert}
\newcommand{\be}{\begin{equation}}
\newcommand{\ee}{\end{equation}}
\def\ie{{\it i.\ e.\ }}
\def\trace{{\rm tr}\;}
\newcommand{\fnorm}[1] {\vert \vert #1 \vert \vert_F}

\begin{document}
\title{Violation of area-law scaling for the entanglement entropy in spin $1/2$ chains}
\author{G. Vitagliano, A. Riera and J. I. Latorre}

\affiliation{
Dept. d'Estructura i Constituents de la Mat\`eria,
Universitat de Barcelona, 647 Diagonal, 08028 Barcelona, Spain
}

\begin{abstract}
Entanglement entropy obeys area law scaling for typical
physical quantum systems. This may naively be argued to follow from
locality of interactions. We show that this is not the case by
constructing an explicit simple spin chain Hamiltonian with nearest neighbor interactions
that presents an entanglement volume scaling law. This non-translational model is contrived to 
have couplings that force the accumulation of singlet bonds across the half chain. 
Our result is complementary to the known relation between non-translational invariant,
nearest neighbor interacting Hamiltonians and QMA complete problems.

\end{abstract}
\maketitle

\section{Introduction}

Ground states of relevant physical Hamiltonians carry 
quantum correlations which decrease with distance. 
For instance, a two-point correlation function is
expected to fall off exponentially with the separation of
points in the presence of a mass gap or algebraically
at critical points. This amount of quantum correlations
is in correspondence with an \emph{area law} scaling of 
entanglement entropy.
To be precise, the entanglement entropy,
defined by
\be
S(\rho_A)=-\trace \left(\rho_A \log \rho_A \right) \, ,
\ee
where $\rho_A$ is the density matrix of the region of space $A$ considered,
scales as the boundary of $A$. 

In contradistinction to the above situation, random quantum states
are known to carry volume law entropy \cite{Page:1993-71}. Therefore, typical
Hamiltonians produce grounds states which are not generic. Indeed,
relevant physics corresponds to a small corner of the total
Hilbert space of a quantum system. This observation is crucial
to understand recent efforts to simulate quantum states with
tensor networks \cite{White:1992-69, Vidal:2003-91, Nishino:2000-575, PerezGarcia:2007-7, Vidal:2008-101, Verstraete:2004-ar, Verstraete:2008-57}. Such approximations are able to accommodate
area law scaling for the entropy.

It is then important to understand precisely what are
the properties that a Hamiltonian must obey so as to produce
a ground state which only displays area law entanglement.
A first heuristic approach suggests that entanglement 
decreases at large distances because interactions are local.
That is, a local degree of freedom interacts with its 
neighbor and gets entangled with it. This second degree of
freedom interacts in turn with a further one. This sequence
of interactions would eventually entangle far separated 
degrees of freedom, though the strength of interactions
would only manage to get the standard correlations we
find in Nature. On the other hand, it is unclear whether
interactions could be contrived to achieved larger
entangled states. The role of translational invariance
is then critical.

Some results for one-dimensional systems are well-established.
In one dimension, if a system obeys local interactions and it is gapped, 
area-law always emerges \cite{Hastings:2007-08}.
On the other hand, if the system is at a critical point, and therefore gapless,
a logarithmic divergence is encountered. 
This logarithmic scaling of the entanglement entropy is explained by
conformal field theory \cite{Vidal:2003-90, Latorre:2004kg, Calabrese:2004-0406,Calabrese:2009-42, Latorre:2009-42}.

In Refs.\ \cite{Wolf:2006-96, Farkas:2007-48}, infinite translational invariant 
Fermionic systems of any spatial dimension with arbitrary interactions are considered.
For such systems, it is shown that the entropy of a finite region typically 
scales with the area of the boundary times a logarithmic correction. 

Although there has recently been further progress on this topic
in higher dimensions
 \cite{Riera:2006-74, Amico:2008-80, Eisert:2008-review},
the necessary and sufficient conditions for an area-law have not been defined yet.
An explicit example of a system where area-law is violated is presented in Ref.\ \cite{Aharonov:2009-287}.
It is shown that a one dimensional non-translational invariant system 
composed  of 12-level local quantum particles
with nearest neighbor interactions presents a ground state that carries a volume law scaling
of entanglement.
In particular, it is proven that the problem of approximating 
the ground state energy of such system is QMA-complete.
This precise example shows that a quantum computer could not simulate any one dimensional system,
and, moreover, that there exist one-dimensional systems which take an exponential time to relax 
to their ground states at any temperature, making them candidates for being
one-dimensional spin glasses. 

The issue addressed in this work is to study how simple can be a quantum system 
to give a highly entangled ground state.
In particular,  we show that a simple spin 1/2 model with nearest neighbors interactions
with a suitable fine tunning of its coupling constants can have a ground state
with a volume law scaling for the entanglement entropy.
Our proposal is based on the translational symmetry breaking, 
hence, this makes that the area-law violation can not be maintained 
for any bipartition of the system. Nevertheless, it will be shown that 
the average of the entanglement entropy over all the possible positions of the block fulfills a volume-law. 
Our results are presented in the following way. We first review the real space
renormalization group technique which brings the fundamental intuition on how to build
an area law violating Hamiltonian. We then turn to solve the proposed Hamiltonian
using its exact diagonalization, where the final step can be taken both in perturbation
theory or numerically. We also illustrate the real space renormalization idea in an Appendix.


\section{Real space Renormalization Group}
\subsection{Introduction to real space Renormalization Group approach}
Real-space Renormalization Group (RG) approach was introduced in Ref.\ \cite{Fisher:1994-50} 
generalizing the works presented in Ref.\ \cite{Dasgupta:1980-22}.
It is a method suited for finding the effective low energy Hamiltonian
and the ground state of random spin chains. 
The couplings have to satisfy the hypothesis of strong disorder, \ie
the logarithm of its probability distribution is wide.
Under such conditions, the ground state of the system can be very well approximated
by a product state of singlets whose spins are arbitrarily distant.

Let us review the real-space RG method for the inhomogeneous XX model case
\be
 H_{XX} =
\frac{1}{2}\sum_{i=1}^N J_{i}\left(\sigma_i^x \sigma_{i+1}^x+
 \sigma_i^y \sigma_{i+1}^y\right)\, .
\label{eq:HXX}
\ee
First, we find the strongest bond $J_i\gg J_{i+1}, J_{i-1}$ and 
diagonalize it independently of the rest of the chain. 
According to the previous Hamiltonian, this leads to a singlet between spins
$i$ and $i+1$ (see appendix \ref{appendix:perturbation-theory}). 
Therefore, the ground state at zeroth order in perturbation theory respect the couplings
$J_{i-1}$ and $J_{i+1}$ is
\be
\ket{\psi^{(0)}}=\ket{\psi_{x<i}}\ket{\psi_-}\ket{\psi_{x>i+1}}
\ee
where $\ket{\psi_-}=\frac{1}{\sqrt{2}}\left(\ket{01}_{i,i+1}- \ket{10}_{i,i+1} \right)$
is a singlet state between the spins $i$ and $i+1$,
 and $\ket{\psi_{x<i}}$ and $\ket{\psi_{x>i}}$ correspond to the states
of the rest of the system.

In order to compute corrections to the ground state at higher orders,
we use perturbation theory as it is shown in appendix \ref{appendix:perturbation-theory}.
This leads to an effective interaction
between the distant spins $i-1$ and $i+2$ with an effective coupling
\be
\tilde J_{i-1, i+2}=\frac{J_{i-1}J_{i+1}}{2 J_{i}} \, .
\label{eq:ef-bond}
\ee
In summary, real space RG integrates out two spins and reduces the Hamiltonian energy
scale. Notice that this new effective low energy Hamiltonian couples the spins $i-1$
and $i+2$, therefore, it has non-local interactions as seen from the original Hamiltonian.
Iterating this procedure for a XX model with random couplings, 
it is seen that the ground state can be described by a random singlet phase, \ie
each spin forms a singlet pair with another one (see Fig.\ \ref{fig:singlet-phase}a). 
Most pairs involve nearby spins, but some of them produce long distance correlations.

In Ref.\ \cite{Refael:2004-93}, real-space RG was used to 
show that, for random spin chains where the ground state is a random singlet phase,
the entanglement entropy scales logarithmically
at the critical point as in the homogeneous case. That is,
\be
S_L\sim \frac{\tilde{c}}{3} \log_2 L \, ,
\ee
where $\tilde{c}=c \ln 2$ is an effective central charge
proportional the central charge for the same model but without disorder $c$.
This analytical result has later been check numerically 
in Refs. \cite{Laflorencie:2005-72,Chiara:2006-0603,Binosi:2007-76}.

\subsection{Area-law violation for the entanglement entropy}
\label{sec:our-XX-model}
\begin{figure}
\begin{center}
\begin{tabular}[t]{lr}
(a) & \scalebox{.7}{\includegraphics{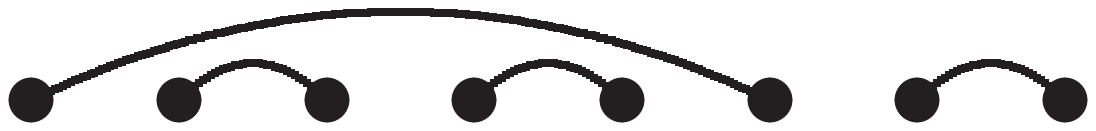}} \\
(b) & \scalebox{.7}{\includegraphics{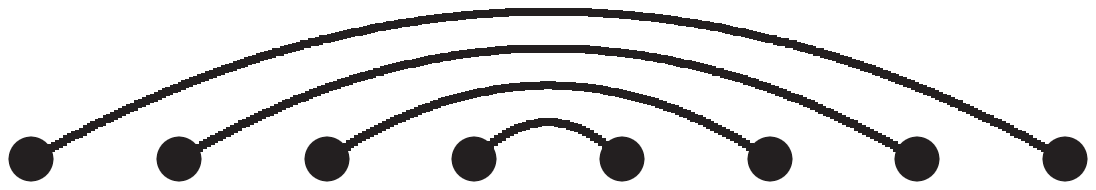}}
\end{tabular}
\end{center}
\caption{
Diagram of a random singlet phase (a) and the concentric singlet phase (b). 
Each spin forms a singlet pair with another spin indicated by the bond lines.
}
\label{fig:singlet-phase}
\end{figure}

\begin{figure}
\begin{flushleft}(a) \\ 
\end{flushleft}
\begin{center}
\scalebox{.7}{\includegraphics{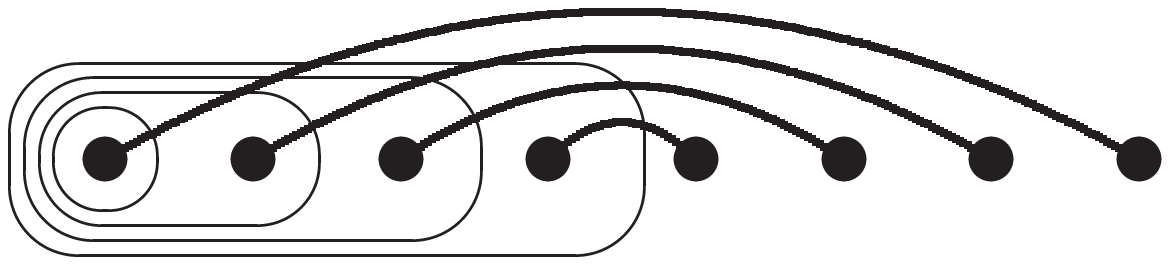}} \\
\end{center}
\begin{flushleft}(b) \\ 
\end{flushleft}
\begin{center}
\scalebox{.7}{\includegraphics{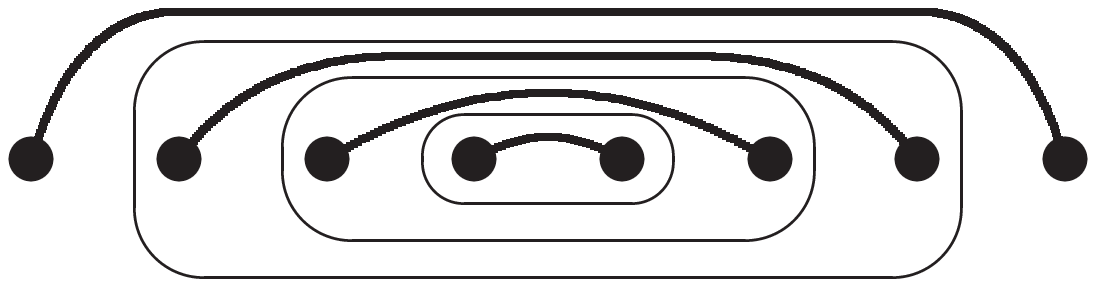}}
\end{center}
\caption{
Diagram of the entanglement entropy scaling for the
concentric singlet phase. The entanglement entropy grows maximally if we take blocks at one extreme (a)
and is zero if the blocks are centered at the middle of the chain.
This is an explicit example that in the non-translationally invariant 
systems the entanglement entropy depends on
the position of the block.
}
\label{fig:block-position}
\end{figure}

Let us now tune the couplings $J_i$ of our XX model in such a way that 
the entanglement entropy of the ground state of the system scales with the volume of the
block of spins. 
An easy way of achieving this is to generate a ground state with
a concentric singlet phase as it is shown in Fig.\ \ref{fig:singlet-phase}b.
We see that the system is in a product state of distant singlets between the positions
$N/2-(i-1)$ and $N/2+i$ for $1\le i\le N/2$.
It is trivial to see that the entanglement entropy of this configuration would
scale with the size of the block, since it merely corresponds to the
number of bonds cut by the bipartition (see Fig.\ \ref{fig:block-position}a). 

Let us note that
the entanglement entropy for concentric blocks would be 0 as it is shown in
Fig.\ \ref{fig:block-position}b.
As translational invariance of the system is broken, the entanglement entropy of 
a block not only depends on the size of it but also in its position.

In order to measure how entangled is a state for non-translationally invariant
systems, it is useful to introduce the average 
entanglement entropy over all the possible positions of the block, that is 
\be
\bar S_L=\frac{1}{N-L}\sum_{i=1}^{N-L}S_L(i) \, 
\ee
where $S_L(i)$ is the entanglement entropy of the block of size $L$ from the $i$-th spin
to the $(i+L)$-th one.

According to the previous definition, 
the average entanglement entropy of the concentric singlet phase
reads
\be
\bar S_L=\left( 1- \frac{L}{2(N-L)}\right)L \, .
\ee
Although for the concentric singlet phase the average entropy losses its 
linear behavior for large blocks, $L \sim \left(1-\frac{1}{\sqrt 3} \right)N$, 
it always fulfills the condition $\bar S_L\ge \frac{1}{2}L$. 
Thus, the concentric singlet phase represents
a simple and explicit example of area-law violation of scaling of the entanglement.

The aim of our work is to tune the coupling constants $J_i$ of the XX model, such
that, the concentric singlet phase becomes the ground state of the system, and, 
in this way, to obtain an explicit example of a Hamiltonian with nearest neighbor interactions
of spins that violate the area-law scaling of entanglement.

Due to the symmetry of the state that we pretend to generate, 
let us consider a XX chain of $N$ spins where the central coupling between
spins $N/2$ and $N/2+1$ is $J_0$ and the rest of them are chosen as follows
\be
J_{N/2+i,N/2+i+1}=J_{N/2-i,N/2-i+1}\equiv J_i  \, ,
\ee
where $1\le i \le \frac{N}{2}-1$ and the coupling $J_{N/2\pm i}$ connects the
spins $N/2\pm i$ and $N/2\pm i+1$.

We are going to use real space renormalization group ideas in order to
see at which values we have to tune the coupling constants, such that, 
the concentric singlet phase becomes the ground state of the system.
If $J_0\gg J_1$, in the low-energy limit, an effective interaction between
the spins $N/2-1$ and $N/2+2$ appears. 
We label this effective coupling as $\tilde J_1$ and, 
according to Eq.\ (\ref{eq:ef-bond}), it reads
\be
\tilde J_1=\frac{J_1^2}{2 J_0} \, .
\ee
Then, if $\tilde J_1\gg J_2$, the effective low-energy Hamiltonian will have an effective bond
between the spins $N/2-2$ and $N/2+3$.
We would like to proceed in this way in order 
to generate iteratively the concentric singlet phase.

Thus, if the condition $\tilde J_i\gg J_{i+1}$ is fulfilled in general, where $\tilde J_i$
is defined by
\be
\tilde J_i=\frac{J_i^2}{2 \tilde J_{i-1}} \, ,
\label{eq:ef-bond2}
\ee
we expect that the ground state of the system is the concentric singlet phase.
\begin{figure}
\begin{center}
\scalebox{1}{\includegraphics{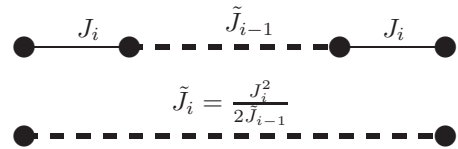}}
\end{center}
\caption{
Diagram of the formation of an effective coupling $\tilde J_i$ if 
the condition $\tilde J_{i-1}\gg J_i$ is fulfilled.
}
\label{fig:effective-bond}
\end{figure}

Specifically, if we impose that $J_i=\epsilon \tilde J_{i-1}$ for any $i$, such that
it is always possible to apply Eq.\ (\ref{eq:ef-bond2}), we see that the couplings
$J_i$ must decay very rapidly
\be
J_i=\epsilon \left(\frac{\epsilon^2}{2} \right)^{i-1}J_0 \, .
\label{eq:bonds-decayA}
\ee
In general, we are going to study chains with couplings that decay
\be
J_i=\epsilon^{\alpha(i)} \, ,
\label{eq:bonds-decayB}
\ee
where $\alpha(i)$ is a function that is monotonically increasing. 
If $\alpha(i)\sim i^2$, we would have a Gaussian decaying. 

Next, we want to solve the XX model with the coupling constants 
defined in Eq.\ (\ref{eq:bonds-decayB}), and
study how the entanglement entropy scales depending on
the kind of decay law for the couplings.

\section{Solution of a spin model and its entanglement entropy}
Let us consider a finite spin chain with 
nearest neighbor couplings $J_{i}^x$, $J_{i}^y$  
and an arbitrary transverse magnetic field $\lambda_i$ in each spin. 
This system is described by the Hamiltonian:
\be
 H =-\frac{1}{2}\sum_{i=1}^N 
 \left(J_{i}^{x}\sigma_i^x \sigma_{i+1}^x+J_{i}^{y}\sigma_i^y \sigma_{i+1}^y\right)
 -\sum_{i=1}^N \lambda_i \sigma_i^z
 \label{eq:H}
\ee
where $L$ is the size of the system and $\sigma_i^{x,z}$ are Pauli-matrices  at site $i$. 
The XX model presented before is a particular 
case of this Hamiltonian (\ref{eq:H}) for
$J_{i}^{x}=J_{i}^{y}$ and $\lambda_i=0$ $\forall i$.
 

\subsection{Jordan-Wigner transformation}
The essential technique in the solution of $H$ is the well-known mapping to
spinless fermions by means of the Jordan-Wigner transformation. 
First, we express the spin operators 
$\sigma_i^{x,y,z}$ in terms of fermion creation (annihilation) operators 
$c_i^\dagger$ ($c_i$):  
$c^\dagger_i=a_i^+\exp\left[\pi i \sum_{j}^{i-1}a_j^+a_j^-\right]$
and $c_i=\exp\left[\pi i \sum_{j}^{i-1}a_j^+a_j^-\right]a_i^-$, 
where $a_j^{\pm}=(\sigma_j^x \pm i\sigma_j^y)/2$. 
Doing this, $H$ can be rewritten in a quadratic form
in fermion operators
\be
H=\sum_{i,j=1}^N A_{ij}c^\dagger_ic_j+\frac{1}{2}\sum_{i,j=1}^N B_{ij}
\left(c^\dagger_ic^\dagger_j+h.c.\right)\ ,
\label{eq:fermions}
\ee
where the matrices $A$ and $B$ are defined by
\begin{align}
A_{ij}&=2\lambda_i \delta_{i,j}+(J_{i}^x +J_{i}^y)\delta_{i+1,j}+(J_{j}^x +J_{j}^y)\delta_{i,j+1} \nonumber \\
B_{ij}&=(J_{i}^x -J_{i}^y)\delta_{i+1,j}-(J_{j}^x -J_{j}^y)\delta_{i,j+1} \, ,
\label{eq:A-B-definition}
\end{align}	
with $1\le i,j \le N$.

\subsection{Bogoliubov transformation}
In a second step, the Hamiltonian is diagonalized using a Bogoliubov transformation
\be
\eta_k=\sum_{i=1}^N\left( \frac{1}{2}\left(\Phi_k(i)+\Psi_k(i)\right)c_i+
   \frac{1}{2}\left(\Phi_k(i)-\Psi_k(i)\right)c_i^\dagger\right)
\label{eq:Bogoliubov}
\ee
where the $\Phi_k$ and $\Psi_k$ are real and normalized vectors: $\sum_i^N
\Phi_k^2(i)=\sum_i^N \Psi^2_k(i)=1$, so that we have
\be
H=\sum_{k=1}^N \Lambda_k(\eta_k^\dagger \eta_k-1/2).
\label{eq:free_fermion}
\ee
The fermionic excitation energies, $\Lambda_k$, and the components of the
vectors, $\Phi_k$ and $\Psi_k$, are obtained from the
solution of the following equations:
\begin{align}
\label{eq:Phi-Psi}
(A-B)\Phi_k&=\Lambda_k \Psi_k \\
(A+B)\Psi_k&=\Lambda_k \Phi_k \, .
\label{eq:Psi-Phi}
\end{align}
It is easy to transform them into an eigenvalue problem,
\begin{align}
\label{eq:eigenvalue-problemI}
(A+B)(A-B)\Phi_k&=\Lambda_k^2 \Phi_k \\
(A-B)(A+B)\Psi_k&=\Lambda_k^2 \Psi_k \, ,
\end{align}
from where $\Lambda_k$, $\Phi_k$ and $\Psi_k$ can be determined.

\subsection{Ground State}
In Eqs. (\ref{eq:Psi-Phi}) and (\ref{eq:Phi-Psi}), 
we realize that transforming 
$\Phi_k$ into $-\Phi_k$ (or $\Psi_k$ into $-\Psi_k$), 
$\Lambda_k$ is changed to $-\Lambda_k$.  
This allows us to restrict ourselves to the sector corresponding to
$\Lambda_k \ge 0$, $k=1,2,\dots,N$. 
Thus, considering Eq.\ (\ref{eq:free_fermion}) and the fact that all $\Lambda_k$
are positive, the ground state is a state $|GS\rangle$ which verifies,
\be
\eta_k |GS\rangle = 0  \ \ \ \ \forall k \ .
\ee

In practice, what we do to restrict ourselves to the sector of positive $\Lambda_k$ is to
determine $\Phi_k$ and $\Lambda_k$ by solving Eq.\ (\ref{eq:eigenvalue-problemI}), 
and calculate $\Psi_k=\frac{1}{\Lambda_k}(A-B)\Phi_k$.

\subsection{Computation of  Von Neumann entropy }
Following Refs.\ \cite{Chung:2001-64, Peschel:2003-36,Latorre:2004kg}, 
the reduced density matrix $\rho_{L}=\trace_{N-L} |GS\rangle \langle GS |$
of the ground state of a block of $L$ sites in a system of free fermions can be written as
\be
\rho_L = \kappa e^{- \mathcal{H}} \, ,
\ee
where $\kappa$ is a normalization constant and $\mathcal{H}$ a free fermion Hamiltonian.

Let us very briefly justify why the density matrix must have this structure.
First, notice that the Hamiltonian defined by Eq.\ (\ref{eq:fermions}) 
has Slater determinants as eigenstates.
Thus, according to Wick theorem, any correlation function of the ground state
(or any other eigenstate)
can be expressed in terms of correlators of couples of creation and annihilation operators. 
For instance,
\be 
\langle c_n^\dagger c_m^\dagger c_k c_l \rangle =
\langle c_n^\dagger c_l \rangle \langle c_m^\dagger c_k \rangle 
-\langle c_n^\dagger c_k \rangle \langle c_m^\dagger c_l \rangle
+\langle c_n^\dagger c_m^\dagger \rangle \langle c_k c_l \rangle\, .
\ee
If all these indices belong to a subsystem of $L$ sites, 
the reduced density matrix $\rho_L$ must reproduce the expectation
values of the correlation functions, \ie
\begin{align}
\trace \left(\rho_L c_n^\dagger c_m^\dagger c_k c_l\right)&=
\trace \left(\rho_L c_n^\dagger c_l \right)\trace \left(\rho_L c_m^\dagger c_k\right)\nonumber \\
&-\trace \left(\rho_L c_n^\dagger c_k\right)\trace \left(\rho_L c_m^\dagger c_l \right)  \\
&+\trace \left(\rho_L c_n^\dagger c_m^\dagger\right)\trace \left(\rho_L c_k c_l\right) \nonumber \, .
\end{align}
This is only possible if $\rho_L$ is the exponential of 
an operator $\mathcal{H}$ which also contains creation and annihilation processes, \ie
\be
\mathcal{H}=\sum_{i,j=1}^L \tilde A_{ij}c^\dagger_ic_j+\frac{1}{2}\sum_{i,j=1}^L \tilde B_{ij}
\left(c^\dagger_ic^\dagger_j-h.c.\right)\, . 
\label{eq:Hdensity-matrix}
\ee

We can diagonalize this Hamiltonian $\mathcal{H}$ by means of another Bogoliubov transformation 
\be
\xi_k=\sum_{i=1}^L\left( \frac{1}{2}\left(v_k(i)+u_k(i)\right)c_i+
   \frac{1}{2}\left(v_k(i)-u_k(i)\right)c_i^\dagger\right)\, ,
\label{eq:Bogoliubov2}
\ee
where $v_k(i)$ and $u_k(i)$ are real and normalized.
Then, the Hamiltonian reads
\be
\mathcal{H}= \sum_{k=1}^{L} \epsilon_k \xi^\dagger_k \xi_k \, ,
\ee
where $\xi^\dagger_k$ and $\xi_k$ are the creation and annihilation operators of some fermionic modes.
In terms of these modes, the density matrix $\rho_L$ is uncorrelated and can simply be expressed as 
\be
\rho_L=\otimes^L_{k=1}\tilde \rho_k \, 
\ee
where
\be
\tilde \rho_k =
\frac{1}{1+e^{-\epsilon_k}}
\left(\begin{array}{cc}
e^{-\epsilon_k} & 0 \\
0 & 1
\end{array}\right) 
=
\left(\begin{array}{cc}
\frac{1+\nu_k}{2} & 0 \\
0 & \frac{1-\nu_k}{2}
\end{array}\right) \, .
\ee
In the previous equation, the new parameters $\nu_k$ have been introduced in
order to ensure the normalization of $\tilde \rho_k$, $\trace \left(\tilde \rho_k \right)=1$.
This way of expressing $\tilde \rho_k$ will be useful next.

Thus, the entanglement entropy of the density matrix
$\rho_L$ is merely the sum of
binary entropies
\begin{equation}
\label{eq:entanglement-entropy}
S(L)=\sum_{k=1}^L S(\tilde \rho_k) 
  =\sum_{k=1}^L H\left(\frac{1+\nu_k}{2}\right) \, .
\end{equation}
where $H(p)\equiv -p\log_2 p -(1-p)\log_2(1-p)$ is
the binary Shannon entropy.

In order to determine the spectrum of $\tilde\rho_k$, 
let us consider the correlation matrix,
\be
G_{m,n}\equiv\langle GS\vert (c_n^\dagger- c_n)(c_m^\dagger+c_m)\vert GS \rangle \, .
\label{eq:correlation-matrix}
\ee
Notice that the matrix $G$ can be computed using the $\Phi_k$ and $\Psi_k$ vectors,
\be
G_{m,n}=-\sum_{k=1}^N \Psi_k(m)\Phi_k(n),
\label{eq:correlation-matrix2}
\ee
where the correlations $\langle \eta_k^\dagger \eta_q\rangle=\delta_{kq}$ and 
$\langle \eta_k \eta_q\rangle=0$ have been considered.

In the subspace of $L$ spins, $G$ is completely determined 
by the reduced density matrix.
To avoid any confusion, let us define $T\equiv G(1:L,1:L)$ 
as the $L\times L$ upper-left sub-matrix
of the correlation matrix $G$. 
Then, $T$ can be expressed in terms of the expected values $\langle \xi_k^\dagger \xi_q \rangle$,
\begin{align}
T_{i,j}&=\sum_{k,q=1}^L u_k(i) v_q(j)  
  \left( \langle \xi_k^\dagger \xi_q \rangle - \langle \xi_k \xi_q^\dagger \rangle \right) \nonumber \\
&= \sum_{k=1}^L u_k(i) v_k(j) \nu_k \, ,
\end{align}
where the $i$ and $j$ indices run from 1 to $L$.
This equation leads to the relations,
\begin{align}
T u_q &= \nu_q v_q \\
T^T v_q &= \nu_q u_q \, ,
\end{align}
that can be translated to the eigenvalue problem
\begin{align}
T^T T u_q &= \nu_q^2 u_q \\
T T^T v_q &= \nu_q^2 v_q \, .
\label{eq:eigenproblemII}
\end{align}
Once the $\nu_q$ variables are computed, we can determine the entanglement entropy
by means of Eq.\ (\ref{eq:entanglement-entropy}).

\subsection{Summary of the calculation}
\label{sec:steps-calculation}
To sum up, let us enumerate the steps that we have to follow
in order to calculate the entanglement entropy of a block $L$.
\begin{enumerate}
\item Write down the matrices $A$ and $B$ in terms of the couplings of the Hamiltonian (\ref{eq:H})
according to Eqs.\ (\ref{eq:A-B-definition}).
\item Determine $\Lambda_k$, $\Phi_k$ and $\Psi_k$ 
by solving the eigenvalue problem from Eq.\ (\ref{eq:eigenvalue-problemI}).
\item Calculate the correlation matrix $G$ defined in Eq.\ (\ref{eq:correlation-matrix}).
\item Take the sub-matrix $T$ and 
      to determine the eigenvalues $\nu_k$ from Eq.\ (\ref{eq:eigenproblemII}).
\item Compute the entanglement entropy by means of Eq.\ (\ref{eq:entanglement-entropy}).
\end{enumerate}

\section{Expansion of the entanglement entropy}
We would like to tune the coupling constants of the Hamiltonian (\ref{eq:H}),
such that the scaling of the entanglement entropy of its ground state
violates the area law.
The entanglement entropy only depends on the variables $\nu_k$.
Then, we can separate the Shannon entropy of the probabilities 
$\frac{1\pm\nu_k}{2}$ into 
\be
H\left(\frac{1+\nu_k}{2}\right)=1-h(\nu_k)\, ,
\ee
where $h(x)=-\frac{1}{2}\log(1-x^2)-\frac{x}{2}\left(\frac{1-x}{1+x}\right)$, 
is a positive function.
Thus, the entanglement entropy reads
\be
S(L)=L -\sum_{k=1}^L h(\nu_k) \, .
\ee
Notice that the scaling of the entropy only depends
on the sum $\sum_{k=1}^L h(\nu_k)$. 
More concretely, we can define the parameter, 
\be
\beta \equiv \lim_{L\to\infty}\frac{1}{L}\sum_{k=1}^L h(\nu_k)\, , 
\ee
that describes
the asymptotic behavior of the scaling of the entropy for large blocks:
\begin{itemize}
\item $\beta = 0$: maximal entanglement,
\item $\beta < 1$: volume-law,  
\item $\beta = 1$: sub-volume-law.
\end{itemize}
Let us focus on the case $\beta\sim0$. Let us analyze if it is possible
to design a spin chain with nearest neighbor interactions whose
ground state is maximally entangled.
First, we realize that $\beta$ is strictly zero if and only if
all the variables $\nu_k=0$. 
Thus, if we want to consider small deviations of the
maximally entangled case, 
we can assume that $\nu_k\sim 0$ and expand $\beta$ in series
of $\nu_k$,
\be
\beta =\lim_{L\to \infty} \frac{1}{2L} \sum_{k=1}^L \left( \nu_k^2 + O(\nu_k^4) \right)\, .
\ee
Considering Eq.\ (\ref{eq:eigenproblemII}), we can express $\beta$ in terms of
the matrix-elements of $T$,
\be
\beta=\lim_{L\to \infty}\trace \left(T T^T\right)=\lim_{L\to \infty}\sum_{i,j=1}^LT_{ij}^2=0\, .
\label{eq:condition-T}
\ee
Let us notice that to fulfill this condition requires that the average of the matrix-elements of $T$ 
tend to zero for large $L$,
\be
\lim_{L\to \infty} \frac{1}{L^2} \sum_{i,j=1}^L \vert T_{i,j} \vert=0 \, .
\ee
If we assume a smooth behavior for the matrix-elements of $T$, 
according to Eq.\ (\ref{eq:condition-T}), they must decay faster than 
the inverse square root function,
\be
T_{ij}\sim\frac{1}{(ij)^{\frac{1}{2}+\epsilon}} \, ,
\ee
such that $\beta=0$. 

In conclusion, in order that the entanglement entropy scales
close to the maximal way, the matrix-elements of $T$ matrix
have to be very close (or decay rapidly) to zero.
If this is the case, the entanglement entropy can be simplified to
\be
S(L)=L -\fnorm{T}^2 \, ,
\label{eq:entropy-approximation}
\ee
where $\fnorm{T}$ is the Frobenius norm
of $T$, defined by 
$\fnorm{T}=\sqrt{\trace\left( T^T T\right)}$.

Let us now study if it is possible to tune the coupling constants of
a spin chain in order that $\fnorm{T}$ is strictly (or close to) zero.
The possibility of having a null $T$ is discarded because it cannot be achieved
with nearest neighbor interactions models.
Despite this, there is a wide freedom to tune the
coupling constants such that the matrix-elements of $T$ fulfill
condition (\ref{eq:condition-T}). 
This arbitrariness makes very difficult to specify the shape of
the distribution of coupling constants in order that area-law is
violated.
With this aim, we can exploit the idea of real space Renormalization 
Group  presented before.

\section{Numerical Results}
We can follow the steps described in Sec.\ \ref{sec:steps-calculation} in order to 
calculate the entanglement entropy of the XX chain presented in Sec. \ref{sec:our-XX-model}
and check if the entanglement entropy grows linearly with the size of the block.

This XX model is characterized by having the strongest bond in the middle of the chain, $J_0$, 
while the value of the rest of bonds $J_n$ decrease rapidly with the distance $n$ to 
the central one. 
In particular, we have studied two different kinds of decay for the coupling constants $J_n$: 
({\it i}) \emph{Gaussian} decay, $J_n=e^{-n^2}$, and
({\it ii}) \emph{exponential} decay $J_n=e^{-n}$. 
Let us notice that due to the rapidly decaying of the coupling constants and
the finite precision of the computer, 
we can only consider small systems.

\begin{figure}
\begin{center}
(a)
\scalebox{.7}{\includegraphics{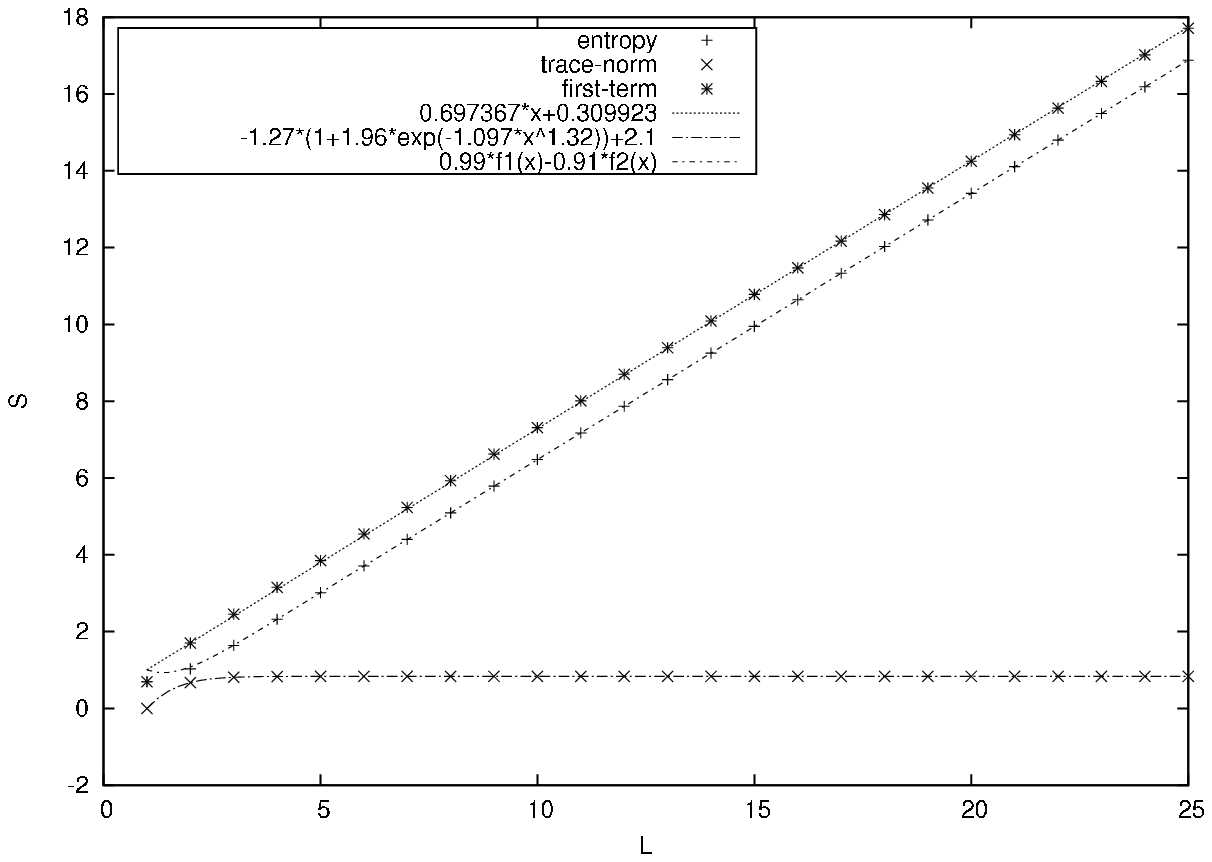}}
(b)
\scalebox{.7}{\includegraphics{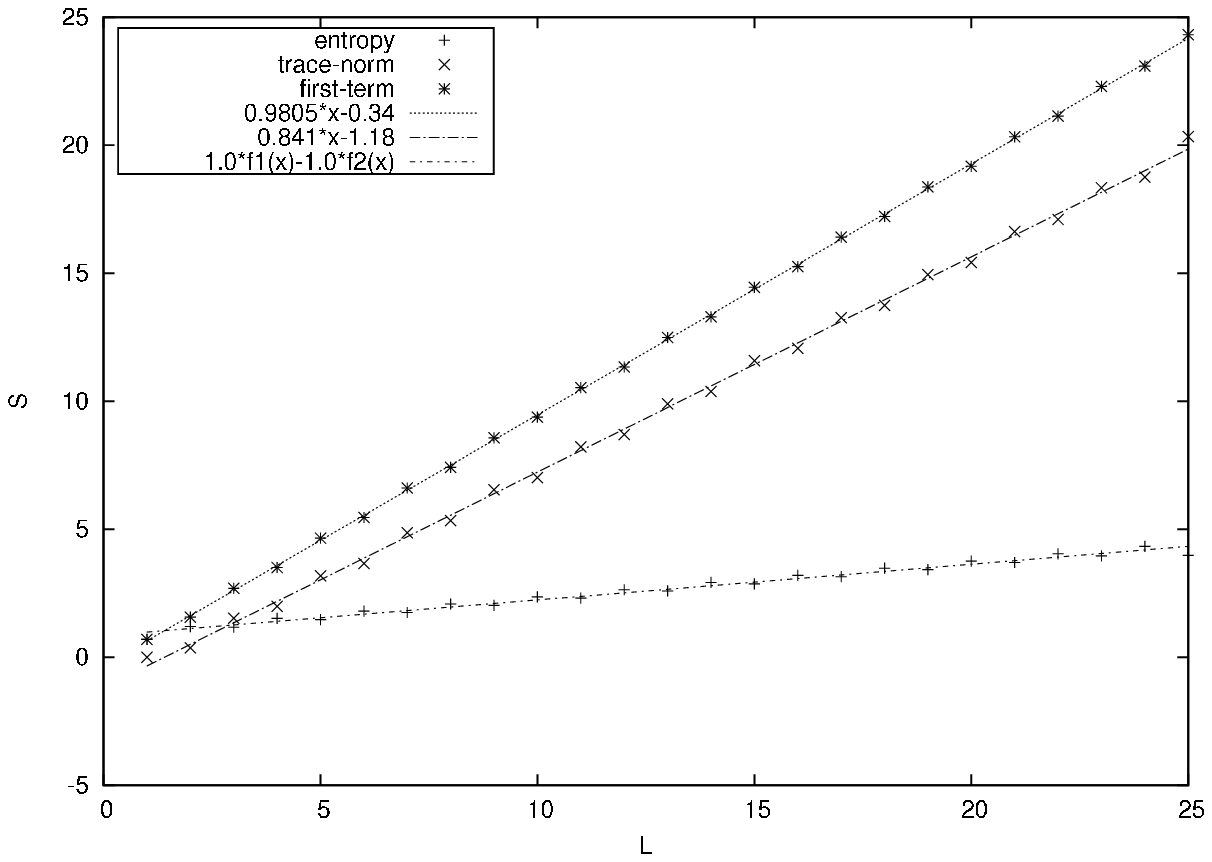}}
\end{center}
\caption{
Scaling of the entanglement entropy of a block of
contiguous spins with respect to the size of the block $L$
for the ground state of a XX model with couplings 
that decay: (a) in a Gaussian, $J_n=e^{-n^2}$,
and (b) in an exponential, $J_n=e^{-n}$, way. 
The magnetic field is set to zero.
The Frobenius norm of the $T$ matrix related to this system
and the sum $S(L) + \fnorm{T}$ are also plotted.
This allows us to check how accurate is the approximation
(\ref{eq:entropy-approximation})
 both in the case of Gaussian decay (a) and in the
exponential one (b).
}
\label{fig:entropy-scaling}
\end{figure}

In Fig.\ \ref{fig:entropy-scaling}(a), the entanglement entropy is plotted 
for the Gaussian case.
As we expected scales linearly with the size of the block $L$ with a slope
practically equal to one. 
Notice that although the slope is 1 for large blocks,
the entanglement is not the maximal due to the non-linear behavior
of the entropy for the smaller ones.
This can be better understood analyzing if the 
approximation of the previous section 
given by Eq.\ (\ref{eq:entropy-approximation}) is fulfilled. 
With this aim, 
the square Frobenius norm $\fnorm{T}^2$ and the sum $S(L) + \fnorm{T}^2$
are also plotted. 
In fact, we observe that the sum $S(L) + \fnorm{T}^2$ coincides with the
maximal entropy, as Eq.\ (\ref{eq:entropy-approximation}) suggests.

The same plot can be realized for an exponential 
decay of the coupling constants, see Fig.\ \ref{fig:entropy-scaling}(b).
In this case, although the entanglement entropy also scales linearly,
its slope is less than one. Thus, we observe a volume law, but the
entropy is not maximal.
Therefore, Eq.\  (\ref{eq:entropy-approximation}) is not fulfilled in this case.
We can, actually, see that the Frobenius norm $\fnorm{T}$
 increases linearly with $L$ instead
of saturating to a small value.

We have repeated the same computations for the same kind of decays
but other basis.
The same behaviors for the Gaussian and the exponential cases have
been obtained. 
For the Gaussian case, a faster decay implies a saturation to a smaller value for 
$\fnorm{T}$, that is, a closer situation to the maximal entropy.
For the exponential case, $\fnorm{T}$ continues increasing linearly
but with a smaller slope.


\section{Conclusions}

We have constructed a one dimensional system composed by spin-$\frac{1}{2}$ particles with 
nearest neighbor interactions with a entanglement entropy of the ground state
that scales with the volume of the size of the block. 

This results further confirms that violations of area law scaling for entanglement
entropy are possible for local interacting Hamiltonians. Furthermore, such a
behavior is found possible for spin 1/2 degrees of freedom.
The price to be paid for violating area law scaling is to break the translational symmetry of the 
system. Indeed, in Refs. \cite{Fannes:2003-44, Farkas:2005-46}, 
it is shown that, although translation-invariant one dimensional states give
rise to arbitrary fast sublinear entropy growth, they cannot support a linear scaling.

Let us also recall that two-local Hamiltonian problems have been proven to
provide QMA-complete problems. To be more precise, the problem of finding out
whether the ground state energy of a two-local Hamiltonian is larger than $a$
or smaller than $b$, where $|a-b|>{\cal O}(1/n)$ is QMA-complete. We may further
argue that an efficient classical simulation of such a problem is likely to
be impossible, otherwise we could solve any NP-complete problem by just simulating
Quantum Mechanics on a classical computer. The obstruction to obtain faithful
simulations of quantum mechanical systems is in turn related to the amount of
entanglement found in the system. Thus, exponentially large entanglement
should be found in some one-dimensional quantum systems. Our results is
somehow completing this idea. Even spin 1/2 chains can produce highly
entangled states if couplings are adequately tuned.

\appendix

\section{Real space renormalization group in a XX model of 4 spins}
\label{appendix:perturbation-theory}
We consider first a simple XX model with only 4 spins and couplings $\{\lambda, 1, \lambda \}$.
We can rewrite the Hamiltonian of the system as a perturbation theory problem,
\begin{equation}
H=H_0 + \lambda V \, ,
\end{equation}
where,
\begin{equation}
H_0= \sigma^X_2 \sigma^X_3 +\sigma^Y_2 \sigma^Y_3 \, ,
\end{equation}
and
\begin{equation}
V= \sigma^X_1 \sigma^X_2 +\sigma^Y_1 \sigma^Y_2  + \sigma^X_3 \sigma^X_4 +\sigma^Y_3 \sigma^Y_4\, .
\end{equation}
The eigenstates of $H_0$ are
\begin{align}
\ket{\psi_+}&=\frac{1}{\sqrt{2}}\left(\ket{01}_{23}+ \ket{10}_{23} \right) \nonumber \\
\ket{\psi_0}&=\ket{00}_{23} \nonumber \\
\ket{\psi_1}&=\ket{11}_{23} \\
\ket{\psi_-}&=\frac{1}{\sqrt{2}}\left(\ket{01}_{23}- \ket{10}_{23} \right) \nonumber
\end{align}
with eigenvalues +2, 0, 0 and -2 respectively. 
We are interested in study what happens to the ground state (GS) of the Hamiltonian $H$ 
when the perturbation $\lambda V$ is introduced. 
The ground state of $H_0$ is degenerate and form a subspace of dimension 4. 
In particular, we choose the set of vectors 
$\{\ket{m}\}=$ $\{\ket{0}_1\ket{\psi_-}_{23}\ket{0}_4$, $\ket{0}_1\ket{\psi_-}_{23}\ket{1}_4$,
$\ket{1}_1\ket{\psi_-}_{23}\ket{0}_4$, $\ket{1}_1\ket{\psi_-}_{23}\ket{1}_4 \}$ as a basis.


We expect that the perturbation removes the degeneracy in the sense that 
there will be 4 perturbed eigenstates all with
different energies. 
Let us call them $\{ \ket{l} \}$. 
As $\lambda$ goes to zero, $\ket{l}$ tend to $\ket{l^{(0)}}$ 
which are eigenstates of $H_0$, but which in general will not coincide with $\ket{m}$.

According to perturbation theory, let us expand the eigenstates and the eigenvalues of $H$ in powers of $\lambda$, 
\begin{equation}
\ket{l}=\ket{l^{(0)}}+\lambda \ket{l^{(1)}}+\lambda^2 \ket{l^{(2)}}+O(\lambda^3)
\end{equation}
and 
\begin{equation}
E_l=E_{GS}^{(0)}+\lambda E_l^{(1)}+\lambda^2 E_l^{(2)}+O(\lambda^3) \, .
\end{equation}
Notice that the zero order term in the energy expansion does not depend on $l$, 
since the ground state of the non-perturbed Hamiltonian is degenerate.
Substituting the previous expansions into the Schr\" odinger equation, 
$\left(H_0 + \lambda V \right)\ket{l}= E_l \ket{l}$,
and equating the coefficient of various powers of $\lambda$, we obtain a set of equations that will allow us to 
find the corrections to the perturbed eigenstates and eigenvalues.

At zero order in $\lambda$ we recover the trivial non-perturbed Schr\" odinger equation. 
If we collect terms of order $\lambda$, we get
\begin{equation}
\label{1st-order}
\left(E_D^{0}-H_0\right)\ket{l^{(1)}}=\left(V-E_l^{(1)}\right)\ket{l^{(0)}} \, .
\end{equation}
In order to calculate the first correction to the energy, 
we project the previous equation (\ref{1st-order}) to the degenerate ground state subspace
\begin{equation}
\sum_{m'=1}^4 V_{m, m'} \langle m' \ket{l^{(0)}} = E^{(1)}_l \langle m\ket{l^{(0)}} \, ,
\end{equation}
where $V_{m, m'}\equiv \bra{m}V \ket{m'}$ is the projection of the interaction to this subspace.
In our particular case, the matrix-elements $V_{m, m'}=0$ for all $m$ and $m'$, hence, $E^{(1)}_l=0$ $\forall$ $l$. 
This means that the degeneration is not broken at first order in $\lambda$ 
and forces us to consider the second order,
\begin{equation}
\label{2nd-order}
\left(E_D^{0}-H_0\right)\ket{l^{(2)}}=\left(V-E_l^{(1)}\right)\ket{l^{(1)}} - E_l^{(2)}\ket{l^{(0)}}\, .
\end{equation}
We proceed as previously and project this equation to the degenerate ground state subspace,
\begin{equation}
\label{2nd-order-B}
\bra{m} V-E_l^{(1)} \ket{l^{(1)}} = E^{(2)}_l  \langle m\ket{l^{(0)}} \, .
\end{equation}
From equation (\ref{1st-order}) we can compute the first order correction to the eigenstates $\ket{l}$,
\begin{equation}
\ket{l^{(1)}}=\sum_{k \notin GS} \frac{ \bra{k^{(0)}}V\ket{l^{(0)}} } { E_{GS}^{(0)}-E_k^{(0)} }
\end{equation}
where $\ket{k^{(0)}}$ are the $H_0$ eigenstates that do not belong to GS. 
Now, we substitute this into (\ref{2nd-order}) and get an equation  for the second order correction to the energies and
the states $\ket{l^{(0)}}$
\begin{equation}
\label{eigenproblem}
\sum_{m',k} \frac{ \bra{m} V \ket{k^{(0)}} \bra{k^{(0)}} V \ket{m} }{ E_{GS}^{(0)}-E_k^{(0)} } \alpha^l_m
=E_l^{(2)} \alpha^l_m
\end{equation}
where $\alpha^l_m$ are the coefficients of $\ket{l^{(0)}}\equiv\sum_m \alpha^l_m \ket{m} $ expressed in terms of the basis $\ket{m}$. Notice that Eq.\ (\ref{eigenproblem}) is a diagonalization problem. 
For our particular Hamiltonian, it takes the form
\begin{equation}
2
\begin{pmatrix}
1 & 0 & 0 & 0 \\
0 & 1 & 1 & 0 \\
0 & 1 & 1 & 0 \\
0 & 0 & 0 & 1 
\end{pmatrix}
\cdot
\begin{pmatrix}
\alpha_1  \\  \alpha_2  \\ \alpha_3  \\ \alpha_4  
\end{pmatrix}
=
E_l^{(2)}
\begin{pmatrix}
\alpha_1  \\  \alpha_2  \\ \alpha_3  \\ \alpha_4  
\end{pmatrix}
\, .
\end{equation}
Now the degeneration is completely broken and the perturbed ground state becomes
\begin{equation}
\ket{GS}=\ket{\psi_-}_{14}\ket{\psi_-}_{23}-\lambda \frac{1}{\sqrt{2}} (\ket{1001}+\ket{0110}) +O(\lambda^2) ,
\end{equation}
with 
\begin{equation}
E_{GS}=-2 + O(\lambda^3)\, .
\end{equation}
We can obtain an effective Hamiltonian by projecting the original one 
into the subspace of lower energy formed by $\{\ket{l^{0}}\}$,
\be
H_{eff}=P H P^\dagger = -2 + \frac{\lambda^2}{2}\left(2+\sigma^X_1 \sigma^X_4 +\sigma^Y_1 \sigma^Y_4\right) \, ,
\label{eq:Heff}
\ee
where $P=\sum_{l^{(0)}}\ket{l^{(0)}}\bra{l^{(0)}}$ and 
$\ket{l^{0}}\in \{ \ket{\psi_-}_{14}\ket{\psi_-}_{23}$,  
$\ket{\psi_0}_{14}\ket{\psi_-}_{23}$, $\ket{\psi_1}_{14}\ket{\psi_-}_{23}$,
$\ket{\psi_+}_{14}\ket{\psi_-}_{23}\}$.
\begin{acknowledgments}
We thank F. Verstraete, M. Asorey, J. G. Esteve and P. Calabrese for fruitful
discussions. 
Financial support from QAP (EU), MICINN (Spain), FI
program and Grup Consolidat (Generalitat de Catalunya), and
QOIT Consolider-Ingenio 2010 is acknowledged.
\end{acknowledgments}

\bibliographystyle{apsrev} 
\bibliography{arealaw-violation}

\end{document}